# Lamarckian inheritance following sensorimotor training and its neural basis in *Drosophila*

Ziv M Williams [1-3]

[1.] Harvard-MIT Health Sciences and Technology, Boston, MA
[2.] Harvard Medical School Program in Neuroscience, Boston, MA
[3.] Department of Neurosurgery, MGH-HMS Center for Nervous System Repair, Harvard Medical School, Boston MA

Contact: zwilliams@mgh.harvard.edu

**Jean-Baptiste Lamarck was among first to suggest that certain acquired traits may be heritable from parents to offspring. In this study, I examine whether and what aspects of sensorimotor conditioning by parents prior to conception may influence the behavior of subsequent generations in *Drosophila*. Using genetic and anatomic techniques, I find that both first- and second-generation offspring of parents who underwent prolonged olfactory training over multiple days displayed a distinct response bias to the same specific trained odors. The offspring displayed an enhanced anemotactic approach response to the trained odors, however, and did not differentiate between orders based on whether parental training was aversive or appetitive. Consequently, disruption of both olfactory-receptor and dorsal-paired-medial neuron input into the mushroom bodies abolished this change in offspring response, but disrupting synaptic output from α/β neurons of the mushroom body themselves had little effect on behavior even though they remained necessary for enacting newly trained conditioned responses. These observations identify a unique transgenerational dissociation between parentally-trained conditioned and unconditioned sensory stimuli, and provide a putative neural basis for how sensorimotor experiences in insects may bias the behavior of subsequent generations.**

In the late 1700's, Jean-Baptiste Lamarck suggested that certain acquired traits may be inheritable [1,2]. Since then, extensive studies have demonstrated that parental experience can indeed have a profound effect on offspring phenotype [3-7]. Behaviorally, both adult and *in utero* exposure to factors such as malnutrition or stress can also influence behavioral traits such as anxiety [8-13]. Similarly traumatic parental experience can lead to certain behavioral changes, such as anxiety or hyperactivity, in subsequent generations [10,14-16].

Whether and by what neural process specific *acquired* traits are transmittable from parents to offspring, however, remains poorly understood. For instance, by repeated conditioning, animals can rapidly learn to associate a sensory cue, such as an odor, with an unconditioned stimulus such as painful shock in order to produce a specific conditioned response such as avoidance [17,18]. Several factors, however, are thought to limit the potential transmission of such acquired information across generations. Most notably, learned associations are encoded by neural ensembles within the brain which have no clear way of translating task-specific information to an animal's progeny [19]. Moreover, there has been little direct evidence in mammals to suggest that prior sensorimotor learning in parents leads to similar conditioned behavior in their offspring. Recently, however, it has been shown that mice trained on odors paired with an electric shock sire offspring that display heightened startle response to the same odors [20]. This suggests that certain acquired traits may potentially be transmittable from parents to offspring prior to conception.

Here, I aimed to examine whether and what unique information about conditioned sensory cues, unconditioned stimuli and conditioned responses may be inheritable when parents are subject to prolonged sensorimotor training. Second, I aimed to examine what neural processes, at the circuit-level, may be causally involved in mediating such an effect. By using anatomic and genetic techniques in a *Drosophila* conditioning model, I find that both first- and second-generation offspring of parents who underwent prolonged olfactory training displayed a distinct response bias to the same trained sensory stimuli, displaying an enhanced anemotactic response. These responses were not observed when subjecting the parents to the odors or unconditioned stimulus alone or when temporally uncoupling their presentation. At the circuit level, both genetic and anatomic disruption of sensory input into the mushroom bodies of the offspring abolished this selective change in behavior, but disrupting synaptic output from the mushroom bodies themselves had little effect. These findings suggest that salient sensory stimuli in the environment can significantly influence the behavior of subsequent insect generations and provides a basic neural roadmap by which to understand this transgenerational process. This effect was not due to direct induction and was odor-specific.

# MATERIALS & METHODS

**Experimental setup**

The goal of the study was to examine whether and what aspects of prolonged sensorimotor training in adult flies (*Drosophila melanogaster*) influences offspring behavior, and define the neural circuit mechanism by which this may occur. Unlike prior studies that have largely focused on short-term memory using brief training periods over a single generation, I focuses here on the prolong F0 training over multiple consecutive days and followed its effect across generations. Towards this end, groups of adult *Drosophila* underwent either prolonged appetitive or aversive olfactory conditioning, and their offspring were followed up to two filial (F1 and F2) generations (see further below; **Fig. 1A**).

To ensure that there was no exposure of the offspring to the experimental stimuli, the flies were raised in the following manner: Odor-naïve adult *Drosophilae* were first cultured together. Within 12 hours of eclosion, newly emerged F0 male and virgin female flies were sorted and separated. Male and virgin female F0 flies then underwent 5 days of either aversive or appetitive conditioning. Only F0 male and confirmed virgin female flies were reintroduced onto new media following training to allow for fertilization. The initial growth medium was also retrospectively followed for up to two weeks to confirm non-parity (see additional male-only control below). To ensure that there was no contact between the F0 and F1 generations, F0 flies were removed from the medium within 1 day after laying eggs, and only F1 adults from this new medium were used for subsequent testing. All F0 flies were chosen randomly and no flies were excluded from the experiments. F1 and F2 generation flies underwent no training and had no exposure to the experimental stimuli prior to testing. Additional controls included male-only training and non-conditioned exposure (see further below).

Dual-odor discrimination enables one to determine whether a specific conditioned stimulus (CS), such as odor 1 vs. odor 2, had become associated with a specific conditioned response (CR), such as preference vs. aversion, based on which unconditioned stimulus (US) they were originally paired with. F0 flies were trained on 3-octanol (OCT) or 4-methylcyclohexanol (MCH) as the CS and received sucrose or mild electric shock as the US. Following training and after separation from the next fly generation (see above), the flies were tested in a T-maze apparatus.

Here, odor concentrations and flow rates were purposefully set to elicit an odor approach response (rather than aversion; see below). The main reason for this was that odor exposure occurred over prolonged durations and over multiple days. Odors at higher concentrations led to a progressive decline in fly behavior and learning when given over many repeated training events. This also sets the present training task somewhat apart from prior studies that have examined the effect of short-term associative training at higher odor concentrations [18,21]. Here, the goal of the experiments was to determine whether and how prolonged olfactory training within and across days in young adult flies influenced the behavior of subsequent generations (preliminary data demonstrated no transgenerational effect when adults were given only brief single-trial and/or single-day training as traditionally done previously; data not shown). All odor and limb selections were made blindly under computer control (AutoMate). Seven thousand two hundred and ten adult *Drosophilae* were tested in total.

**Fly cultures**

*Culture conditions*

Groups of adult flies underwent either prolonged appetitive or aversive olfactory conditioning, and their offspring were followed up to two filial generations (**Fig. 1A**). All experiments were performed in a separately enclosed room under dark, humidity controlled and sound attenuated conditions. Most experiments were carried out at 25 $^0$C (see below). There was no phenotypic selection, and all F0 flies were divided randomly into groups prior to training. All F1 or F2 flies *per* culture vial were used for each experiment. The wild-type adults were cultured at 25 $^0$C on corn starch medium supplemented with live yeast under standard day-light cycle. They were grown in large (50 cc) conical enclosures and culture media for adult flies was replaced every 2-3 days. F0 training sessions were performed after the flies were at least 2-3 day old (younger flies have been found to demonstrate poorer performances). Most T-maze testing sessions were performed on day 6 for F0 adults and on days 3-6 for F1 and F2 adults. F1 and F2 adults underwent no training and were not exposed to any odor prior to testing. To further ensure no exposure of F2 flies to the odors (i.e., from testing), F1 adults were moved to a new growth medium prior to T-maze testing, with the original culture medium serving to grow the F2 generation larva.

**Training protocols**

*Appetitive conditioning protocol*

Two to three days after eclosion, groups of odor-naïve F0 adult flies were randomly selected to undergo appetitive training. On consecutive days, the flies were first placed in an enclosure without food for 16-18 hours (this has been shown to improve learning performance) [22]. They then underwent appetitive training using the following protocol. The flies were placed in a 50 cc conical enclosure with standard corn meal medium and added sucrose granules. They were allowed to feed on the medium for 6-8 hours while an odor was bubbled into the enclosure for 10 minutes, once every hour.

Flies were given either OCT or MCH. Air at low-flow was first bubbled through distilled water at room temperature and then through a 1:100 mixture of odorant (OCT or MCH) and mineral oil. The air flow was directed into the fly enclosure through silicone tubing. The flies were removed from the apparatus after feeding on the medium and were then placed into an empty enclosure again. This sequence was repeated a total of 3 times over 5 days (this calculation excludes the first day). Different groups of flies underwent either OCT training or MCH training (see additional odor exposure and backwards conditioning controls below).

*Aversive conditioning protocol*

For aversive conditioning, the flies were placed in 15 cc conical tubes circumferentially covered with an electrifiable copper grid. A mesh was placed on each end of the tube to allow for the free flow of the odor (OCT or MCH). For training, the selected odor was paired with mild electric shocks delivered in intervals over 5 minutes for a total of three training sets *per* day.

In each training set, the flies were first given the odor for 30 seconds. Then, pulses of electric current (30 biphasic pulses over 500 ms at 90V) were given every 30 seconds for a total of 5 minutes. At the end of the 5 minutes, the odorant was removed and the flies were allowed to rest for 5 minutes. This sequence of 5 minutes of stimulation followed by 5 minutes of rest was repeated three times *per* day. Each group of flies underwent 5 days of training. Different groups of flies underwent either only OCT training or MCH training.

*Odor exposure and backwards conditioning controls*

Similar to the approach described previously [23], a single odorant was used during F0 training in the present experiments. This was done because it was important to limit the possibility of stimulus generalization, and because it was unknown what affect the temporal order of odor and unconditioned stimulus presentations may have on the offspring's behavior. To control for the possible effect of simple odor exposure or sensitization, different groups of flies underwent appetitive or aversive training and their responses were subsequently evaluated by dual-odor T-maze testing (i.e., by discrimination).

Two additional controls were also employed. In the first, odor-exposure control, F0 flies were exposed to the odors alone using the same training protocol as above but without an unconditioned stimulus. In the second, backwards-conditioning control, the conditioned stimulus was given after the unconditioned stimulus. For appetitive backwards conditioning, the flies were allowed to feed on the medium for 6 hours without an infused odor. Ten minutes after being removed from the medium, they were then given either OCT or MCH for another 6 hours at 10 minute regimens. For aversive backwards conditioning, the flies were given three sets of electric shock for 5 minutes each, as above, but were not given an odor. Ten minutes after completing this regimen, they were then given three additional sets of either OCT or MCH odor infusion.

**T-maze testing and blinding protocol**

Prior to performing the experiments, different concentrations of OCT and MCH were tested and titrated to produce an approximate equal distribution of F0 flies between the two limbs of the T-maze. Flies will normally have a constitutive preferential response to most odors when given alone [24-26]. As noted further below, OCT and MCH at 1:100 dilutions in mineral oil and at the flow rates tested produced a roughly 50:50 distribution of flies between limbs (**Fig. 1B**). Therefore, for most experiments, OCT and MCH were given at equal concentrations of 1:100.

Tested groups consisted of approximately 20-30 flies. For each group, all flies were introduced by gentle suction to the midpoint of the T-maze apparatus. The suction was released and each limb of the T-maze was infused with a unique odor (OCT in one limb and MCH in the other limb) using the same bubbling device employed for conditioning.

The bottom of the T-maze received low out-flow suction. The flies were then allowed to move within the apparatus for 5 minutes. For $UAS\text{-}Shi^{ts1}/+$ and $UAS\text{-}Shi^{ts1}/C739$ flies tested at restructure temperature, the flies were incubated at 32 $^0$C for 30 minutes immediately prior to T-maze testing.

To ensure blinding of which odorant was given *per* limb, a computerized infusion system was used. Four solenoids controlled entry of one of the two possible odors into each limb of the T-maze (AutoMate Scientific). Limb-odor selections were made pseudo-randomly by a Matlab routine (MathWorks) that controlled the solenoids through a NIDAQ I/O interface (National Instruments). After 5 minutes elapsed, the flies were separated and the number in each limb was documented. The limb-odor selections were revealed after the flies were counted and tabulated. This procedure was repeated for each group.

**Learning of new associations in F1 flies**

Learning in F0 flies was assessed with MCH and OCT, as described in the Experimental Procedures. To assess new learning in *F1* flies, the odorant benzaldehyde (BEN; 99% purity; Sigma Aldrich) at dilution of 1:10,000 was used. Since this control assay was aimed at determining whether learning of new associations was affected, T-maze testing was performed immediately after aversive conditioning and only after training for a single 5 minute run (appetitive training requires prolonged sessions over several repetitions to provide consistent results). Here, $UAS\text{-}Shi^{ts1}/+$ and $UAS\text{-}Shi^{ts1}/C739$ flies were incubated under restrictive temperature for 30 minutes immediately before training. They were then allowed to learn the new association for 5 minutes. Immediately after learning, they were shifted back to permissive temperature. Thirty minutes after resting at permissive temperature, they were tested again.

**Sensorimotor assays**

*Chemotaxis assay*

To test for differences in response to the individual odors, a chemotaxis assay was performed. The flies were placed in the same T-maze apparatus, as before. However, here, a diluted odor was given in one limb and air bubbled in mineral oil was given in the other. The flies were allowed to select between limbs for 5 minutes, following which they were separated and counted. Two odor dilutions were used for both OCT and MCH; 1:100 and 1:100,000. A concentration of 1:100,000 was chosen since it produced preferential responses that were significantly above chance but were also within the range of responses observed during dual odor-testing (see Main Text). Odors at lower dilutions led to a largely equal distribution of flies, and were therefore not used. In order to limit sensitization to the odors, each group was tested only once. For *UAS-Shi$^{ts1}$/+*, *UAS-Shi$^{ts1}$*/C739, antenna-less and *amnesiac* fly testing, OCT and MCH where used at dilutions of 1:100.

*Negative geotaxis assay*

A negative geotaxis assay was used to test for normal locomotion in *UAS-Shi$^{ts1}$/+*, *UAS-Shi$^{ts1}$*/C739, antenna-less and *amnesiac* flies. Here, the flies were placed into a 50 cc vertical conical tube and allowed to acclimate for 5 minutes. They were then gently tapped to the bottom. The numbers of flies that reached or surpassed the 5 cm vertical mark line within 5 seconds were counted. *UAS-Shi$^{ts1}$* flies performed this assay under either permissive or restrictive temperatures.

**Transgenic and anatomically modified flies**

*UAS-Shi$^{ts1}$*/C739 flies were generated by crossing C739/C739 homozygotes with *UAS-Shi$^{ts1}$*/*UAS-Shi$^{ts1}$* homozygotes. *UAS-Shi$^{ts1}$/+* flies were generated by crossing *UAS-Shi$^{ts1}$*/*UAS-Shi$^{ts1}$* homozygotes with wild-type flies. *UAS-Shi$^{ts1}$* flies were grown at 20 $^{0}$C. Training and testing at permissive temperatures was carried out at 25 $^{0}$C. Testing at restrictive temperatures was carried out at 32 $^{0}$C. Heat shock treatment was made by placing the flies in a 32 $^{0}$C water bath for 30 minutes prior to testing.

While the recessivity of *amn$^1$* is semi-dominant when it comes to learning [27], this is not observed when using multiple training trials as done here (see Discussion). Here, *amnesiac* flies were *amn$^1$*/*amn$^1$* homozygotes. C739 and *amnesiac* flies were supplied by Bloomington (Bloomington Drosophila Stock Center). *UAS-Shi$^{ts1}$* flies were kindly provided by Josh Dubnau (Cold Spring Harbor).

Finally, antenna-less F0 flies were made by surgically removing their third antennal segment under anesthesia as described previously [28,29]. The maxillary pulps were preserved. Surgery was performed using Teflon coated micro-forceps under surgical magnification. Only flies confirmed to have completely removed antennae were used. The flies were then allowed to recover for 1-2 days. Visibly normal F1 larva cultures were noted after the procedure. This is consistent with prior reports demonstrating that most adult flies will continue to mate after complete antennal removal [29]. F1 generation adults displayed normal antenna morphology and normal sensorimotor responses (see Main Text).

**Statistics**

Performance index (PI) was defined as the number of flies choosing the OCT odor side of the T-maze minus the MCH odor side, divided by the total number of flies and then multiplied by 100. Since there was no *a priori* expectation of how F1/F2 flies would behave following appetitive or aversive training, a positive number was arbitrarily selected to indicate preference for OCT and a negative number indicated a preference for MCH. Chemotactic response PI was defined in the same way but, instead, compared odor vs. air limb selections. Locomotion PI was defined as the numbers of flies that reached or surpassed the 5 cm vertical mark line at 5 seconds divided by the total number of flies.

An independent, unpaired t-test was used to evaluate whether individual fly groups displayed a significant affinity or aversion to the parent-trained odor. Significance threshold was Boneferroni corrected for comparison across two parent-trained odors *per* generation (p<0.025). A two-tailed, paired t-test was used to evaluate whether training led to a *selectivity* in response based on training (p<0.05; i.e., they displayed a differential response to OCT vs. MCH based on which odor the F0 flies were trained on). Therefore, if F1 flies display a preferential response to MCH when their parents underwent appetitive training with MCH *and* F1 flies display a preferential response to OCT when their F0 parents underwent appetitive training with OCT, then the t-statistic would be *positive* and significant. Finally, an N-way analysis of variance (ANOVA) was used to evaluate for differences in the absolute magnitude of response between groups and their interaction (p<0.05). All statistics were given with their degrees of freedom (df) based on the number of groups tested, t-statistic (ts) and significance (p). Graphs were given with their standard error of the mean (s.e.m.).

# RESULTS

### F0 appetitive training biases F1/F2 behavior in an odor-selective manner

F0 flies were first tested for long-term retention after 5 days of training (F0 testing was performed on day 6, approximately 1 day after completion of training). As expected, odor-naïve F0 flies that underwent no training demonstrated no differential response to the two odors (**Fig. 1B**). In comparison, F0 flies that underwent prolonged appetitive training demonstrated a strong preferential response to the trained odors (**Fig. 1C**). When compared across both odors, the effect of training was positive and odor-selective (df = 21, ts = +7.4, $P = 2.5 \times 10^{-07}$) meaning that the F0 flies demonstrated a selective preference towards the trained odors.

Even though F1 and F2 flies underwent no training and had no prior exposure to the stimuli, they displayed a weak but surprisingly selective response to the same parent-trained. Specifically, F1 flies whose parents underwent appetitive training displayed a selective but smaller preferential bias towards the odors experienced by their parents (df

= 19, ts = +4.6, $P = 1.8 \times 10^{-04}$). This differential response to the odors persisted in the F2 generation (df = 23, ts = +3.7, $P = 1.2 \times 10^{-03}$). Therefore, for example, if the parents underwent appetitive training with OCT, their offspring displayed a slight but significant preference to OCT over MCH during T-maze testing. F1 and F2 flies whose parents underwent no training demonstrated no differential response to either tested odor (**Fig. 1B**).

**The effect of F0 training on offspring behavior is dependent on both the CS and US**

What aspects of parental experience influenced the offspring's behavior? Exposure of the parents to the odors alone was insufficient to elicit a similar preferential response to the odors in the subsequent generation. In control experiments, F0 flies were given the same odors and training schedule as before but now without the US. Following odor exposure to either OCT or MCH for 5 days, neither F0 nor subsequent F1 fly generations demonstrated a differential response to the parent-exposed odors (df = 10, ts = +0.15, $P = 0.89$). A separate set of F0 flies were also given training sessions in which the CS was given after the US (i.e., backwards conditioning; **Fig. 2A**). Under these settings, the presence of a US during F0 training led to no change in behavior in the F1 flies (df = 10, ts = -0.51, $P = 0.62$; **Fig. 2B**).

A chemotactic assay was also used to assess for differences in odor acuity (i.e., rather than selectivity) based on presentation of a single odor vs. air infusion. As tested here, at low concentrations, flies demonstrate an anemotactic preference to most presented odors [24-26], meaning that they have a constitutive approach response to most odors. This is in contrast to odor *discrimination* which identifies a differential odor preference. Here, it is observed that F1 flies had a stronger differential response to odorants at a 1:100 dilution compared to the lower 1:100,000 dilution (df = 46, ts = -24.76, $P = 3.1 \times 10^{-28}$; **Fig. 3**, note that here the PI indicates preference rather than discrimination; Methods). Therefore, at the dilutions and flow-rates tested, higher odor concentrations led to a stronger affinity to the limb in which the odor was infused compared to that in which air was infused. However, when comparing the distribution of flies between limbs in either the 1:100 dilution (df = 22, ts = -0.78, $P$ = 0.44) or 1:100,000 dilution (df = 22, ts = -0.93, $P$ = 0.36) groups, there was no significant difference between F1 flies whose parents underwent prior training and those whose parents underwent no training (see Discussion, below, for interpretation).

**Aversive training reveals a dissociation between F0 US and F1/F2 response**

Offspring of parents who underwent aversive training did not display avoidance to the parent-trained odors. As anticipated, following aversive conditioning (**Fig. 1A**), F0 flies strongly avoided the trained odors (df = 22, ts = -6.3, $P$ = 2.5 x$10^{-06}$; **Fig. 1D**). However, when the subsequent F1 generation was tested on the same odors, they did not display an aversive response. Rather, they displayed a weak but constitutive preference to the parent-trained odors (**Fig. 1D**). When compared across both odors, parental training had a positive and odor-selective effect on F1 behavior (df = 31, ts = +4.1, $P$ = 2.9 x$10^{-04}$). These behavioral changes persisted in the F2 generation (df = 18, ts = +4.6, $P$ = 2.0 x$10^{-04}$). Similar to appetitive training, there was no change in behavior in F1 flies after F0 aversive backward conditioning (df = 8, ts = -0.46, $P$ = 0.66; **Fig. 3**). In other words, the flies appeared to display an enhanced anemotactic response or preference to *any* odor trained by the parents rather than a CR (see Discussion for further detail).

When further considered across all tested groups, the offspring displayed similar responses across parent-trained odors (OCT/MCH), learning modalities (appetitive/aversive) and generations (F1/F2; ANOVA, df = 101; $P$ = 0.21, $P$ = 0.98, $P$ = 0.79, respectively); with odor-type accounting for the greatest difference in absolute magnitude of response. These findings, therefore, suggested that even though the offspring displayed a selective response to the parent trained CS, this response was not a CR since they did not differentiate between odors trained by their parents under an appetitive or aversive US.

**Male-only F0 appetitive and aversive training**

*Drosophilae* do not gestate or rear their young as do mammals [30]. Moreover, only offspring of confirmed virgin fly cultures were used in the experiments and non-parity was established across all tested groups. Nonetheless, to further rule out the possibility that F1 larva may have been exposed to the experimental stimuli, only F0 male flies were trained and then allowed to fertilize odor-naïve virgin females. Using the same appetitive (df = 9, ts = +3.32, $P = 8.9 \times 10^{-03}$) and aversive (df = 12, ts = +3.58, $P = 3.8 \times 10^{-03}$) training protocol, F1 adults still demonstrated an odor-selective response.

**MB α/β neuron function and F1 response**

In *Drosophila*, input from the olfactory receptor neurons (ORNs) and dorsal paired medial (DPM) neurons into the mushroom bodies (MB) is essential for the formation and maintenance of new sensorimotor associations [31,32]. Synaptic output from α/β neurons of the MB onto the motor control neuropil, in comparison, is necessary for their subsequent retrieval [21,33,34].

The *Shi$^{ts1}$* transgene, which is a dominant-negative temperature sensitive variant of dynamin, was targeted to MB neurons using the well-characterized C739 enhancer trap line [18,35,36] (see Experimental Procedure and Discussion for further detail on construct selection). Consistent with prior reports [36,37], *UAS-Shi$^{ts1}$*/+ and *UAS-Shi$^{ts1}$*/C739 F0 flies demonstrated preserved chemotactic responses, locomotion and learning under both permissive (25 $^0$C) and restrictive (32 $^0$C) temperatures (**Fig. 4**). Retrieval performance, however, was significantly diminished at restrictive temperatures when testing the *UAS-Shi$^{ts1}$*/C739 F0 flies on day 6 following 5 days of training, thus, confirming that inhibition of α/β neurons output was essential for CS-CR retrieval (appetitive: df = 9, ts = +1.07, $P = 0.31$ and aversive: df = 16, ts = -1.34, $P = 0.20$; **Fig. 5A**). In contrast, *UAS-Shi$^{ts1}$*/C739 F1 offspring continued to display a significant preferential response to the parent-trained odors even under restrictive temperatures. This was true of both F1 flies whose parents underwent appetitive (df = 12, ts = +4.27, $P = 1.0 \times 10^{-03}$) or aversive (df = 16, ts = +5.55, $P = 4.4 \times 10^{-05}$) training. *UAS-Shi$^{ts1}$*/C739 F1 flies still displayed diminished retrieval after learning *new* CS-CR associations using a different odorant. While other enhancer trap lines, such as C309 and OK107, can be additionally used as a positive control [21], such constructs were not necessary because the *UAS-Shi$^{ts1}$*/C739 F1 flies did *not* demonstrate disruption of odor-selectivity at restrictive temperatures.

Given the lack of differential response following selective α/β neuron disruption, I tested whether complete, non-selective MB disruption similarly affected behavioral response by using direct hydroxyurea injection [38]. Here, I focused on aversive training since it provided the strongest and most robust effect. When the bilateral MB of F1 flies were injected with hydroxyurea after learned new CS-CR associations and within 12 hours of larva hatching, there was a significant absence of differential response (df = 5, ts = 0.21, $P$ = 0.84). In comparison, F1 flies continued to display largely preserved preferential response to the parent trained odors following F0 training (df = 5, ts = +4.13, $P$ = 9.0 x10$^{-03}$). Therefore, consistent with the behavioral findings, differential response by F1 flies to the parent-trained odors did not appear to depend on synaptic output from the MB even through MB function (see Discussion for further interpretation).

**ORN and DPM neuron function and F1 response**

What role then did ORN function have on F1 behavior? To address this question, it was necessary to disrupt odor reception over prolonged, week-long training durations. It was also important to disrupt olfaction in F0 but not F1 flies. Towards these ends, an anatomic approach was employed whereby the third antennal segment, containing the ORN, was surgically removed in F0 flies [28,29]. As expected, antenna-less F0 flies demonstrated absent chemotactic response to the odors (**Fig. 4**) as well as severely disrupted CS-CR learning (appetitive: df = 8, ts = -0.31, $P$ = 0.76 and aversive: df = 10, ts = +0.02, $P$ = 0.98; **Fig. 5B**). The following generation F1 flies, which now had intact and functioning antenna, demonstrated normal learning when tested on new odors (df = 8, ts = -12.9, $P$ = 1.3x10$^{-07}$). However, F1 flies of antenna-less parents displayed no differential response to the same parent-trained odors. This was true both of F1 flies whose parents underwent appetitive (df = 10, ts = +1.52, $P$ = 0.16) or aversive (df = 10, ts

= -0.07, $P$ = 0.94) training. Therefore, intact parental ORN function was essential for the subsequent change in F1 behavior.

Expression of the *amnesiac* gene leads to constitutively disrupted DPM neuron function and, therefore, disruption of putative US-related input onto the vertical and horizontal lobes (principally α′/β′ neurons) of the MB [22,27,31,32,39]. Moreover, temporally-delayed inhibition of DPM neurons does not affect either learning or long-term CS-CR retrieval (see Discussion for further details on selection of constitutive vs. temporally-controlled inhibition and *amn* rescue).

As expected, $amn^1/amn^1$ F0 flies had largely normal conditioned responses immediately after training (df = 6, ts = -9.10, $P$ = 1.0 x$10^{-04}$), but had poor CS-CR retrieval on day 6 (appetitive: df = 8, ts = +1.25, $P$ = 0.25 and aversive: df = 11, ts = -1.85, $P$ = 0.10; **Fig. 5C**). When the subsequent F1 generations were tested on the same odors, however, they also displayed no differential response based on prior F0 training. This was true of both F1 flies whose parents underwent appetitive (df = 10, ts = +0.43, $P$ = 0.68) or aversive (df = 14, ts = +0.18, $P$ = 0.86) conditioning. Even though the $amn^1/amn^1$ F1 flies did not display a differential response to the parent-trained odors, they importantly continued to display preserved conditioned responses immediately after learning when trained on a new odorant (df = 7, ts = -4.69, $P$ = 2.2x$10^{-03}$). These findings suggested that preserved DPM neuron function was similarly essential for the observed change in F1 behavior.

## DISCUSSION

The present experiments reveal an association between parental training and offspring behavior in *Drosophila* but also reveal a unique and previously unrecognized dissociation between parent-trained CS/US and offspring behavior. Prior human and animal studies, for example, have shown that traumatic parental experience can lead to certain behavioral changes, such as anxiety or hyperactivity, in subsequent generations [10,14-16]. However, these findings are distinct from those made here which reveal a sensory- but not response-selective effect of F0 training on F1/F2 behavior. A more recent remarkable study, in comparison, has shown that traumatic parental experience can lead to a heightened startle responses in mice when presented with the same parent-trained odors [23]. These findings are, therefore, consistent with those made here by demonstrating changes in response to the same parent-trained odors in mice, and a persistence of effect for up to two filial generations. However, they do not reveal what specific aspects of the CS, US and CR influence offspring behavior and do not demonstrate which neural circuitries were causally necessary for the change in offspring behavior (see further below).

In order to distinguish between behavioral changes related to the CS from those related to the CR, flies in the present study were made to select between two simultaneously presented odors rather than being presented with odors individually. Thus, for example, whereas F0 flies that underwent aversive training with OCT preferentially selected MCH over OCT, their F1 offspring selected OCT over MCH. Second, as with many fly conditioning experiments [22,40], the present study was based on differences in group behavior which may not reflect graded differences within individuals. For instance, during the single-odor chemotaxis assay, flies were observed to select one limb of the T-maze over the other and rarely vacillated between limbs during air vs. odor infusions (although certain studies in honeybees have shown that it may be possible to elicit vacillations by thermotaxis when placed at different threshold distances [41]. This also highlights some of the differences between insect and mammal behavior.

Evidence for change in F1 responses in the absence of exposure to the parent-trained odors, use of male-only training and demonstration of persistent differential behavior across two filial generations suggest a biologic inheritance. One common property of the appetitive and aversive training tasks is that they affect neuronally-mediated humoral systems that manifest in widespread physiological changes within an animal [42]. While such effects have been associated with certain epigenetic changes in an animal's gametes [3,7,9,23,43], most such changes are also lost following fertilization [19,44], and these mechanisms do not explain how sensory-related information may be targeted to an animal's gametes. Here, the goal of the study was to identify whether and what specific aspects of sensorimotor learning by parents affect offspring behavior and to determine the neuronal mechanisms responsible for this effect.

A principal benefit of using *Drosophila melanogaster* as an animal model is the ability to modulate specific neurons within the fly's brain, reversibly and across many individuals. *UAS-Shi$^{ts1}$*/C739 flies have been used to temporally control synaptic transmission from the MB and, therefore, disrupt the principal circuit involved in CS-CR retrieval. The sensorimotor performance of *UAS-Shi$^{ts1}$*/C739 flies has also been well-documented and the effect of restrictive temperatures on CS-CR retrieval has been shown to be robust [36,37]. Other enhancer trap lines, such as C309 and OK107, have been used as a positive control to determine whether behavioral changes in constructs such as *UAS-Shi$^{ts1}$*/C747 can be explained by affected circuitries outside the vertical and horizontal lobes of the MB [21]. Use of such constructs, however, was not necessary here because *UAS-Shi$^{ts1}$*/C739 F1 flies did *not* demonstrate disruption of odor-selectivity at restrictive temperatures.

An anatomic approach was further used to constitutively disrupt ORN function. This approach, rather than a transgenic approach was used because F0 training was long and required a consistent way for inhibiting olfaction. For appetitive training, in particular, the flies were conditioned over hours at a time and across multiple days. Transient, temperature-controlled disruption in transgenic lines is normally possible for only brief, non-repetitive training periods [35]. There were two additional benefits to this anatomic approach that made it suitable to the present study. First, antenna removal disrupts olfaction across all odorants and near-completely abolishes CS-CR learning. Second, antennal removal disrupts olfaction only in F0 but not F1 flies, since the latter displayed normal antennae anatomy and chemotactic responses, as shown here.

Finally, *amnesiac* flies were used to study the effect of DPM neuron disruption. The specific *amn$^1$* transgene was employed here because *amn$^1$*/*amn$^1$* homozygotes demonstrate largely preserved learning and short-term retrieval (especially after multiple training trials, as used here) compared to other transgenic lines [27]. Inhibition of DPM neurons, in turn, disrupts the transition of newly formed associations from short- to medium-term memory following F0 training. Formal rescue experiments or other temperature-controllable constructs, such as the C316 enhancer trap-line, have been previously employed to disrupt DPM neuron function in a temporally-delayed fashion. While this can allow for the selective disruption of DPM neurons at later intervals (i.e. many hours after training), such delayed disruption does *not* to affect long-term retrieval [31,32,45] and, therefore, were not necessary here. Moreover, in the present study, no new learning was elicited in the F1 flies and, therefore, delayed disruption was not necessary here. Finally, as noted above, it was critical to maintain consistent DPM neuron disruption over many hours and days, which is generally not possible to achieve with temperature-controllable constructs. These considerations, therefore, made use of the $amn^1/amn^1$ flies optimal for these particular experiments.

Additional investigation may reveal what role neural elements other than those tested here, such as the lateral protocerebrum which is thought to affect experience-independent behavior [46,47], play in enacting these observed responses. Moreover, blockade of MB neuron synaptic output here did not differentially affect F1 responses following either F0 appetitive or aversive conditioning. However, it may be possible that inhibition of surface vs. core α/β neurons or dopaminergic vs. octopaminergic input onto the MB [48,49] could differentially affect avoidance rather than approach behavior.

Taken together, the present study demonstrates an important dissociation between parental sensorimotor training and offspring behavior in insects and explains the neural basis for this effect. Specifically, these observations indicate that adult flies that undergo extended olfactory training have offspring that display selective preferential responses towards the odors experienced by their parents compared to odors that were not. The motor responses themselves, however, are not conditioned responses since the F1 flies do not differentiate between odors that were originally trained by their parents under an aversive vs. appetitive conditions (**Fig. 6**) [22,40,50].

These findings are consistent with the following putative model. Most odors produce a known anemotactic approach response in insects through the ORN [24-26] but do not lead to discrimination or biased preference towards one odor or another when tested at the appropriate concentrations. In F0 flies, olfactory training biases the insect's response either towards or away from the CS and is mediated through the MB. In contrasts, the MB in F1 flies is unaffected by F0 training and odor exposure alone or backward conditioning by the F0 flies do not lead to a change in offspring behavior. Therefore, the observed F1 response to the F0-trained odors were simply an enhanced anemotactic approach response to the F0 odors rather than a CR, leading F1 flies to constitutively prefer to *any* odors previously trained by their parents. Consistently, they were also dependent on intact ORN and DPM function during F0 training [22,27,31,32,39].

This unique transgenerational pseudo-adaptation may provide advantage to animals by allowing parents to prime subsequent generations to attend to salient sensory cues within their environment. Even without information about the specific appetitive or aversive valence of the stimuli, such adaptation may nonetheless allow insects to hone in on sensory cues that were originally deemed important by their parents.

## ACKNOWLEDGEMENTS

I thank Josh Dubnau, Keren Haroush and Shaun Patel for their helpful discussion and for providing input on the experimental design. Z.M.W is supported by the Whitehall Foundation, NIH 5R01-HD059852 and the Presidential Early Career Award for Scientists and Engineers.

## CONFLICTS OF INTEREST

The author has no conflict of interest.

# FIGURES

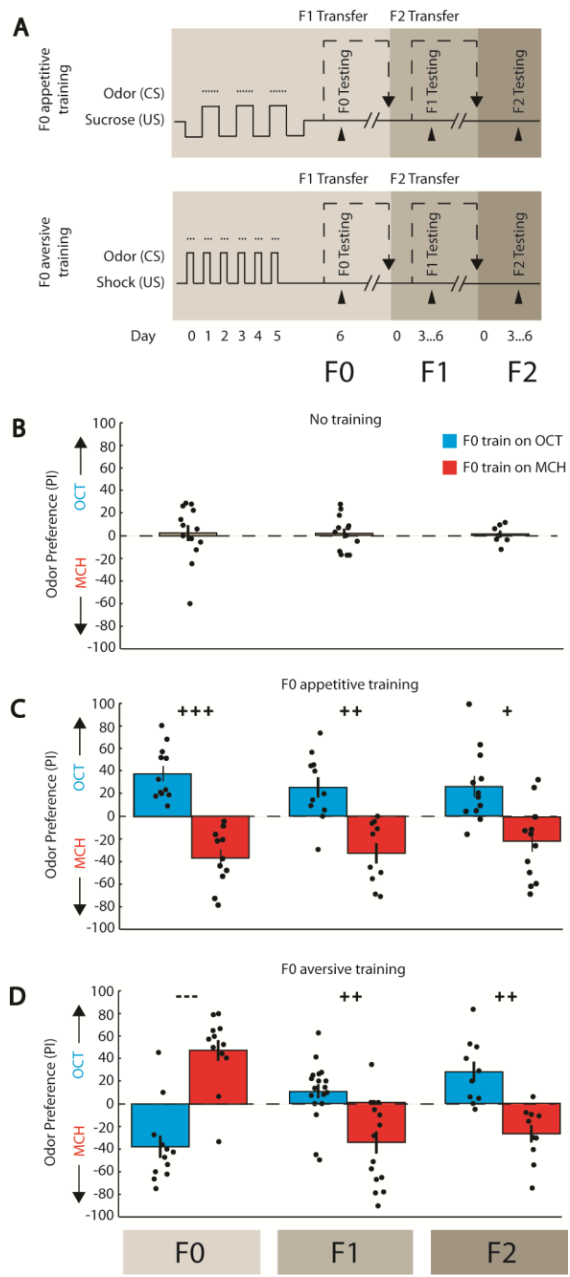

**Figure 1.** Parental olfactory training leads to odor-selective changes in offspring behavior. (**A**) Schematic illustration of the appetitive training protocol for F0 generation flies and subsequent testing of F1 and F2 generation flies. Performance (odor preference) of flies whose F0 parents underwent; (**B**) no training, (**C**) appetitive training or (**D**) aversive training. Bars are color-coded based on whether the F0 generation flies underwent training with OCT (blue) or MCH (red). A positive performance index (PI) indicates a preference towards OCT over MCH and a negative PI indicating a preference to MCH over OCT (Experimental Procedure). Data is presented as the mean ± s.e.m.,

and each dot represents the PI for a single fly group (+++/--- attraction/aversion $P < 0.00001$; ++ attraction, $P < 0.001$; + attraction, $P < 0.01$).

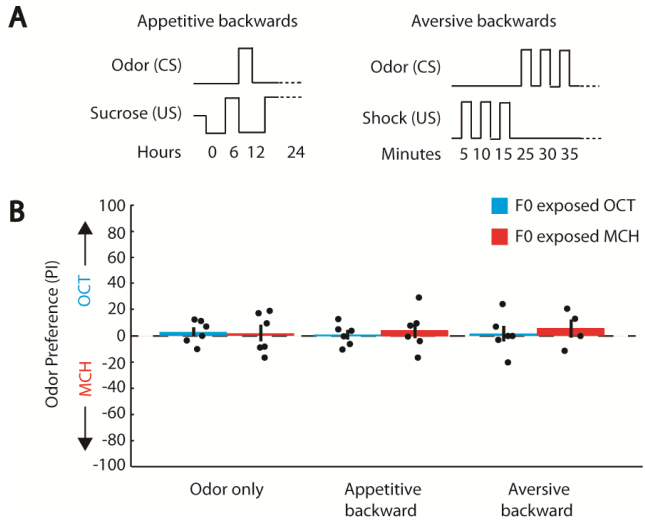

**Figure 2.** Effect of parental odor exposure and backwards conditioning on offspring responses. (**A**) Schematic illustration of the F0 backwards appetitive and aversive conditioning protocols over a single set. (**B**) Performance of F1 flies whose parents underwent odor exposure alone, appetitive backwards conditioning and aversive backwards conditioning. Data is presented as the mean ± s.e.m., and each dot represents the PI for a single fly group.

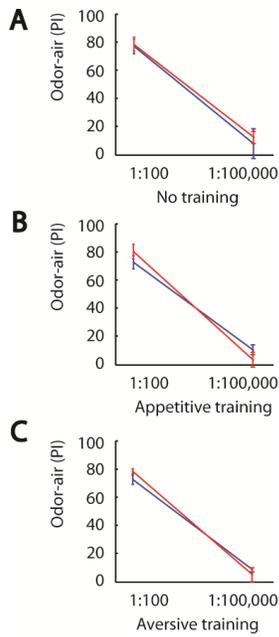

**Figure 3.** Chemotactic responses. Either OCT (blue) or MCH (red) was given in one limb at different concentrations (1:100 or 1:100,000) and air bubbled in mineral oil was given in the other. The PI was defined here as the proportion of flies in the odor limb divided by all flies in both limbs at the end of 5 minutes. Note that when single odors are given alone, flies will display a constitutive preference to the odor and, therefore, the PI in both concentrations here is positive. This metric is distinct from the differential response between opposing odors as shown in **Fig. 1** (see Methods for definition of PI for odor discrimination vs. chemotaxis). Performance of F1 flies whose parents underwent; (**A**) no training, (**B**) appetitive training or (**C**) aversive training. Data is presented as the mean ± s.e.m.

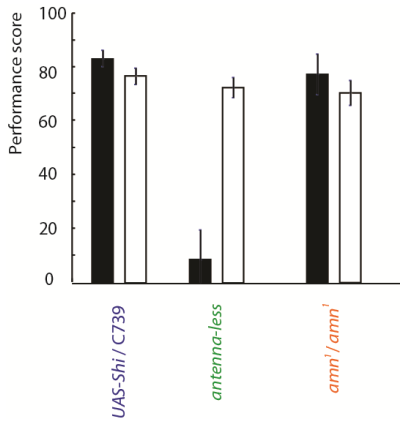

**Figure 4.** Effects of neural disruption on sensory acuity and motor behavior. Effects of UAS-Shi[ts]/C739, restrictive, ORN (antenna-less) and DPM ($amn^1/amn^1$) disruption on F0 chemotactic (black) and geotactic (white) assay performances.

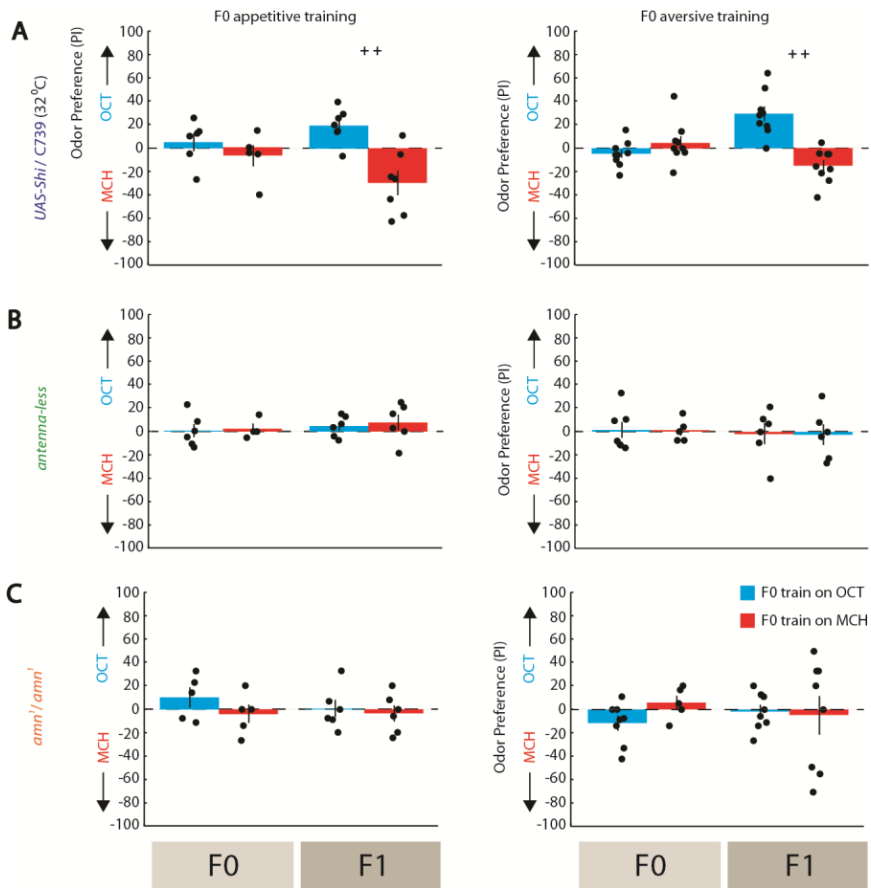

**Figure 5.** Effects of neural disruption on odor-discrimination performance. Effects of (**A**) MB (UAS-Shi$^{ts}$/C739, restrictive), (**B**) ORN (antenna-less) and (**C**) DPM (amn$^1$/amn$^1$) disruption on F0 and F1 odor-discrimination performance following parental appetitive and aversive training. Data is presented as the mean ± s.e.m., and each dot represents the PI for a single fly group (++ attraction, $P < 0.001$).

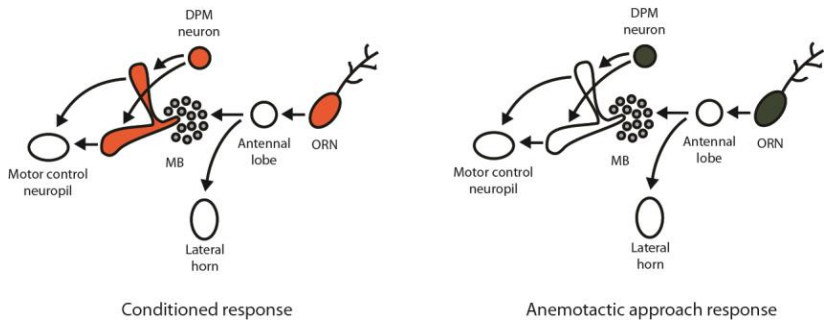

**Figure 6.** Schematic illustration putative circuits involved in F0 and F1 responses following F0 training. Sites necessary for enacting conditioned responses are displayed on the *left* in green whereas sites necessary for enacting anemotactic approach responses are displayed on the *right* in orange.